\documentclass[a4paper,fleqn,usenatbib]{mnras}

\usepackage[T1]{fontenc}
\usepackage{ae,aecompl}

\usepackage{graphicx}	
\usepackage{amsmath}	
\usepackage{amssymb}	

\title[Host Galaxy and Environmental Dependence of AGN]{The Role of Host Galaxy for the Environmental Dependence of Active Nuclei in Local Galaxies}

\author[R.I. Davies et al.]{Richard I. Davies$^{1}$\thanks{E-mail: davies@mpe.mpg.de},
E.K.S.~Hicks$^{2}$,
P.~Erwin$^{1}$,
L.~Burtscher$^{1}$,
A.~Contursi$^{1}$,
\newauthor
R.~Genzel$^{1}$,
A.~Janssen$^{1}$,
M.~Koss$^{3}$,
M.-Y.~Lin$^{1}$,
D.~Lutz$^{1}$,
W.~Maciejewski$^{4}$,
\newauthor
F.~M\"uller-S\'anchez$^{5}$,
G.~Orban~de~Xivry$^{6}$,
C.~Ricci$^{7}$,
R.~Riffel$^{8}$,
R.A.~Riffel$^{9}$,
\newauthor
D.~Rosario$^{10}$,
M.~Schartmann$^{11}$,
A.~Schnorr-M\"uller$^{8}$,
T.~Shimizu$^{1}$,
A.~Sternberg$^{12}$,
\newauthor
E.~Sturm$^{1}$,
T.~Storchi-Bergmann$^{8}$,
L.~Tacconi$^{1}$,
S.~Veilleux$^{13}$
\\
$^{1}$Max-Planck-Institut f\"ur extraterrestrische Physik, Postfach 1312, 85741, Garching, Germany\\
$^{2}$Astronomy Department, University of Alaska, Anchorage, USA\\
$^{3}$Institute for Astronomy, Department of Physics, ETH Zurich, Wolfgang-Pauli-Strasse 27, CH-8093 Zurich, Switzerland\\
$^{4}$Astrophysics Research Institute, Liverpool John Moores University, IC2 Liverpool Science Park, 146 Brownlow Hill, L3 5RF, UK\\
$^{5}$Center for Astrophysics and Space Astronomy, University of Colorado, Boulder, CO 80309-0389, USA\\
$^{6}$Space Sciences, Technologies, and Astrophysics Research Institute, Universit\'e de Li\`ege, 4000 Sart Tilman, Belgium\\
$^{7}$Instituto de Astrof\'isica, Facultad de F\'isica, Pontificia Universidad Cat\'olica de Chile, Casilla 306, Santiago 22, Chile\\
$^{8}$Departamento de Astronomia, Universidade Federal do Rio Grande do Sul, IF, CP 15051, 91501-970 Porto Alegre, RS, Brazil\\
$^{9}$Departamento de F\'isica, Centro de Ci\^encias Naturais e Exatas, Universidade Federal de Santa Maria, 97105-900 Santa Maria, RS, Brazil\\
$^{10}$Department of Physics, Durham University, South Road, Durham, DH1 3LE, UK\\
$^{11}$Centre for Astrophysics and Supercomputing, Swinburne University of Technology, Hawthorn, Victoria, 3122, Australia\\
$^{12}$Raymond and Beverly Sackler School of Physics \& Astronomy, Tel Aviv University, Ramat Aviv 69978, Israel\\
$^{13}$Department of Astronomy and Joint Space-Science Institute, University of Maryland, College Park, MD 20742-2421 USA
}

\date{Accepted XXX. Received YYY; in original form ZZZ}

\pubyear{2016}

\begin{document}
\label{firstpage}
\pagerange{\pageref{firstpage}--\pageref{lastpage}}
\maketitle

\begin{abstract}
We discuss the environment of local hard X-ray selected active galaxies, with reference to two independent group catalogues. We find that the fraction of these AGN in S0 host galaxies decreases strongly as a function of galaxy group size (halo mass) -- which contrasts with the increasing fraction of galaxies of S0 type in denser environments. However, there is no evidence for an environmental dependence of AGN in spiral galaxies. Because most AGN are found in spiral galaxies, this dilutes the signature of environmental dependence for the population as a whole. We argue that the differing results for AGN in disk-dominated and bulge-dominated galaxies is related to the source of the gas fuelling the AGN, and so may also impact the luminosity function, duty cycle, and obscuration. We find that there is a significant difference in the luminosity function for AGN in spiral and S0 galaxies, and tentative evidence for some difference in the fraction of obscured AGN.
\end{abstract}

\begin{keywords}
galaxies: active
-- galaxies: Seyfert
-- galaxies: nuclei
-- galaxies: haloes
-- galaxies: luminosity function
\end{keywords}



\section{Introduction}
\label{sec:intro}

The role of environment in the triggering of active galactic nuclei (AGN) is a topic that has received much attention in the literature.
One reason is that, because we understand that interactions and mergers may lead to gas inflow and hence accretion onto a central massive black hole, there has been a general expectation that AGN should exist in denser environments than inactive galaxies.
However, finding such a link has proven to be difficult and any relations between AGN and environment are weak.
\cite{sab13} summarize some specific issues that may contribute to the lack of a clear consensus.
They point out that there are many definitions of `environment', which may be local or large scale, including a high local density of neighbours, membership of a group or cluster and the relative location within that group or cluster, and galaxy-galaxy interactions (e.g. pairs). 
They also note that the environmental dependence may differ according to AGN luminosity and whether the physical feeding mechanism is radiatively efficient (e.g. X-ray and optical samples) or not (e.g. low excitation radio samples).
An additional complication is the short timescale variability of AGN \citep{hic14}, which implies that comparing AGN to a control sample may completely remove any relation between the AGN and the phenomenon being studied.
As described by \cite{dav14}, a simple way to understand this intuitively is to hypothesize that, for example, gas inflow is triggered at some point in every galaxy that is part of a group. Over a Gyr timescale, inflow will occur at different times in different galaxies, perhaps more than once. In any snapshot, one sees only a random subset of active galaxies, which will be different for another snapshot.
One might conclude from a snapshot that AGN fuelling is related to the group environment; but since at any given time many galaxies in the group are inactive, the conclusion from using a control sample would be that it is not, contradicting the original hypothesis.
A way to avoid this problem could be to look at the incidence of AGN activity as a function of environment, for example as was done by \cite{arn09}.
With a sample of 10 groups and 6 clusters, they find that the fraction of AGN (with $L_{0.3-8keV} > 10^{41}$\,erg\,s$^{-1}$) in groups is a factor 2 higher than in clusters, although the result has marginal significance. Due to concerns about the differing morphological mix of galaxies in these different environments, they also looked at the AGN fraction in early type galaxies, finding a similar result.

Many other studies of the environment of X-ray selected AGN have used correlation analysis to assess the clustering bias on different scales as well as the typical halo mass.
Recent examples include 
\cite{gil09} who found a strong clustering signal for AGN with median $z \sim 0.98$ and $L_{0.5-10keV} \sim 10^{43.8}$\,erg\,s$^{-1}$. The correlation length matched that of galaxies with stellar mass $\ga 10^{10.5}$\,M$_\odot$, and implied a typical halo mass of $\la 10^{12.4}$\,M$_\odot$. 
\cite{fan13} found that AGN ($10^{42} \la L_{2-10keV} [$erg\,s$^{-1}] \la 10^{44}$) typically reside in haloes of mass $\sim10^{13}$\,M$_\odot$.
\cite{dip14} found similar halo masses for the infrared selected AGN they studied at $z \sim 1$, with $10^{13.3}$\,M$_\odot$ and~$10^{12.8}$\,M$_\odot$ depending on whether the AGN are obscured or not.
These papers highlight that the luminosity, selection technique, and redshift of an AGN population are also important factors in assessing environment.
Indeed, \cite{mar13} show that at $z \ga 1$--1.5 there may be a reversal in the incidence of AGN in clusters versus the field.

More locally, using a large number of galaxies from the Sloan Digital Sky Survery (SDSS), with AGN identified via their [O\,III] line emission, \cite{li06} found little difference between the clustering of AGN and control galaxies on scales greater than 1\,Mpc, suggesting that the halos of active and inactive galaxies have similar masses. Between 100\,kpc and 1\,Mpc AGN were more weakly clustered, a result these authors argued could be explained if AGN are preferentially located at the centres of their haloes. And at scales less than 70\,kpc, AGN appeared to be marginally more clustered.
The importance of the local environment was also highlighted by \cite{ser06}, who looked at quasars (identified by $M_i \le -22$) in the SDSS, finding they are overdense on scales $<$100\,kpc, but have no difference compared to the general population at larger 1\,Mpc scales.
This suggests that galaxy-galaxy interactions might be important.
However, from a snapshot survey of AGN (selected in the 0.3-3.5\,keV soft X-ray band, and identified as AGN using optical emission line widths and ratios), \cite{sch00} emphasized the lack of evidence for strong interactions or merging activity.
Indeed, \cite{sab15} have argued that the availability of cold gas in the nuclear regions, rather than local galaxy density or galaxy-galaxy interactions, is the key driver for AGN activity and luminosity.
In contrast, among very hard X-ray selected galaxies the situation may be different, since \cite{kos10} found a higher fraction of interactions and close pairs of galaxies among these AGN than in the general galaxy population.

\cite{sch00} also noted that the host galaxies of AGN appear to be drawn at random from the overall galaxy population, although with a slight bias towards early types.
The difference in the environment for early type and late type AGN hosts was the specific focus of a study by \cite{sou16}, who selected Seyferts from the SDSS via optical line ratios.
They found that the fraction of Seyferts in spiral hosts is independent of their radial location in the halo and the halo mass, a result that matches our findings for spiral galaxies in this paper. However they show that the Seyfert fraction in elliptical galaxies appears to increase at larger cluster radii, although it is generally lower in more massive halos. Reconciling this result with \cite{wil12} who found a clear trend with halo mass for the AGN fraction in elliptical hosts, emphasizes the role of AGN selection. These authors show that in this case the increase in AGN fraction both with stellar mass and with halo mass is driven by radio AGN rather than radiatively efficient Seyferts.

In our analysis here, we employ a technique that is complementary to the commonly used correlation analysis. We use existing group catalogues to assess environment, and look simply at the sizes of the groups (i.e. halo occupation number, which is related to halo mass) in which local, moderately luminous AGN are found.
We make use of the very hard X-ray selected AGN in the {\it Swift BAT} catalogue and look at the environmental dependence separately for the two most common types of host galaxy, spirals and lenticulars.
In doing so, we build on the work of \cite{dav14} and \cite{hic13}, who discuss secular inflow and external accretion in disk-dominated and bulge-dominated hosts.
In addition to their own sample, these authors made use of published integral field spectroscopy data of AGN with matched inactive samples \citep{dum07,wes12}.
Comparing the spatially resolved stellar and gas kinematics, they found that the presence of gas and its co- versus counter- rotation with respect to the stars was consistent with secular inflow from the host galaxy for AGN in spirals but with accretion from the environment for S0 galaxies.
Similar misalignments between the stellar and molecular gas kinematics have also been reported for two early type LINERs \citep{mue13}.
However, these authors could not probe the environment itself.
That is our aim in this paper, to test whether the environmental dependence differs for AGN in spirals versus S0 hosts.

In Sec.~\ref{sec:sample} we describe the sample of AGN we use and the group catalogues with which we cross-correlate them.
The analysis itself is the focus of Sec.~\ref{sec:z0.04}.
Then in Sec.~\ref{sec:lumfunc} and Sec.~\ref{sec:obsc} we discuss the additional topics of luminosity function and obscured fraction, that one may expect to differ if the environmental dependence of AGN activity depends itself on the host galaxy.
We then finish with a summary of our conclusions in Sec.~\ref{sec:conc}.

\section{Sample Selection}
\label{sec:sample}

\subsection{Active Galaxies}
\label{subsec:agn}

\begin{figure*}
\includegraphics[width=12cm]{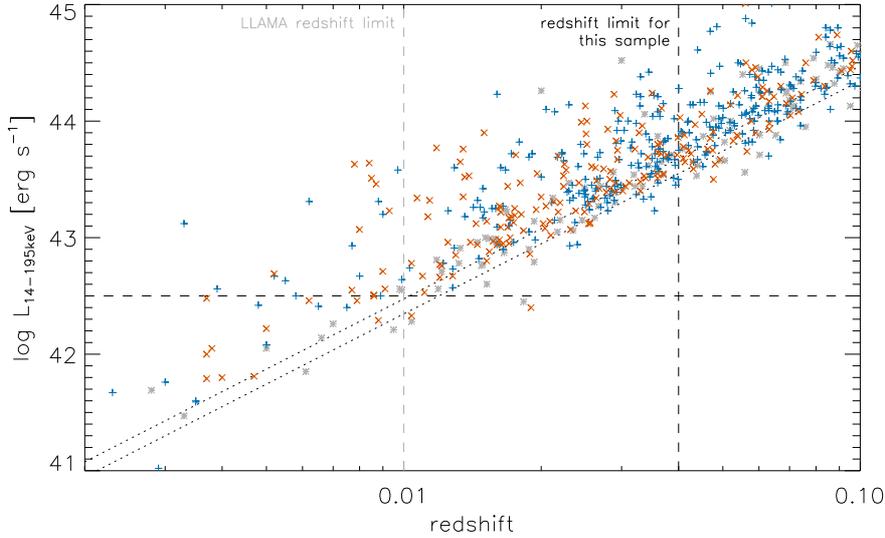}
\caption{AGN in the 70-month {\it Swift BAT} survey (Baumgartner et al. 2013).
Those listed as type~1 are drawn with a blue plus, those listed as type~2 with a red (vermillion) cross, and any AGN without a simple designation as Sy~1 or Sy~2 are marked with a grey asterisk. The redshift limits for the LLAMA sample and the sample analysed here are shown. In both cases the minimum luminosity is $\log{L_{14-195keV}} [$erg\,s$^{-1}] = 42.5$. The diagonal dotted lines indicate the flux limits over 90\% of the sky for the 58-month and 70-month surveys.}
\label{fig:batplot}
\end{figure*}

The sample of AGN we study is related to the LLAMA (Local Luminous AGN with Matched Analogues) sample described in \cite{dav15}, who also give a detailed description of the selection and the rationale for it. 
The key aspect is that these are selected from the all-sky flux limited 14--195\,keV 58-month {\it Swift BAT} survey in such a way as to create a volume limited sample of active galaxies that is as unbiassed as possible, for detailed study using high resolution spectroscopy and adaptive optics integral field spectroscopy.
The sole selection criteria were $z < 0.01$ (corresponding to a distance of $\lesssim40$\,Mpc), $\log{L_{14-195keV}}[$erg\,s$^{-1}] > 42.5$ (using redshift distance), and $\delta < 15^\circ$ so that they are observable from the VLT.
This yielded 20 AGN.
A set of inactive galaxies were selected to match in terms of host galaxy type, mass (using H-band luminosity as a proxy), inclination, presence of a bar, and distance.

Although small, this volume limited sample is sufficient for detailed studies of the molecular and ionised gas kinematics and distributions, as well as the stellar kinematics and populations, in the nuclear and circumnuclear regions.
And, despite being insufficient itself for a statistical analysis of the type discussed in this paper, it provides the rationale for this work.
In such a small sample, the group properties of individual galaxies can be assessed carefully using a variety of catalogues and metrics.
A suprising outcome of doing so was that, while the inactive galaxies (both early and late types) were distributed fairly evenly across the range of environments from clusters through groups to isolated galaxies, the situation was different for the AGN.
While for AGN in late type hosts the environmental distribution matched the inactive late type galaxies, none of the AGN in early type hosts were in large groups or clusters.
This is indicative of a strong environmental effect for AGN in early type hosts, but the number of objects involved is too small to reach a robust conclusion.

In order to increase the number of AGN for the purposes of the study here, we go beyond the LLAMA sample while keeping as close as possible to the original ideal. 
We therefore select all AGN at $z < 0.04$ with $\log{L_{14-195keV}}[$erg\,s$^{-1}] > 42.5$ in the 70-month catalogue \citep{bau13} as shown in Fig.~\ref{fig:batplot}.
In addition we have excluded two objects for which the counterparts were mis-identified and are at redshift higher than our threshold, and included a number of additional sources that meet our criteria as given in Ricci et al. (in prep.) and \cite{ric15}.
This yields 350 AGN.
This sample is no longer volume limited but otherwise imposes no additional bias compared to the LLAMA sample.
An obvious concern would be that the large number of more distant AGN that are excluded because they are below the survey flux limit, may impose a redshift bias on the distribution of host galaxy morphology.
In order to address this, we show in Sec.~\ref{subsec:groupcats} for the specific context of host galaxy morphological classification, that there is no bias with respect to redshift.

\begin{figure}
\includegraphics[width=8cm]{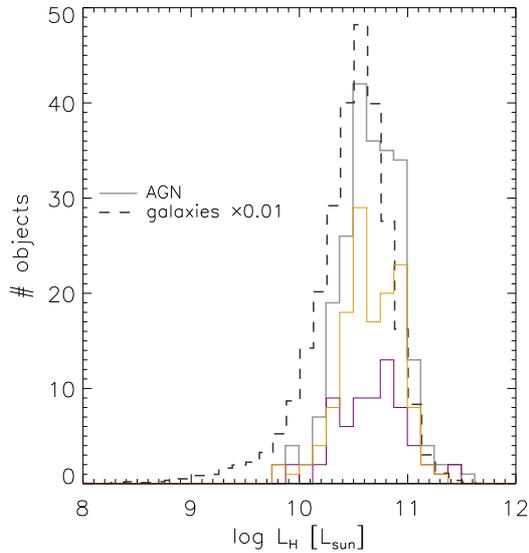}
\caption{Distributions of the H-band luminosity -- as a proxy for stellar mass -- for the AGN sample (solid pale grey line) and comparison galaxy sample (dashed dark grey line) in the Tully (2015) catalogue.
The AGN sample is further split up according to host type, with S0 hosts drawn in purple and spirals in orange.
The distributions are similar.} 
\label{fig:hlum}
\end{figure}

\begin{figure*}
\includegraphics[width=12cm]{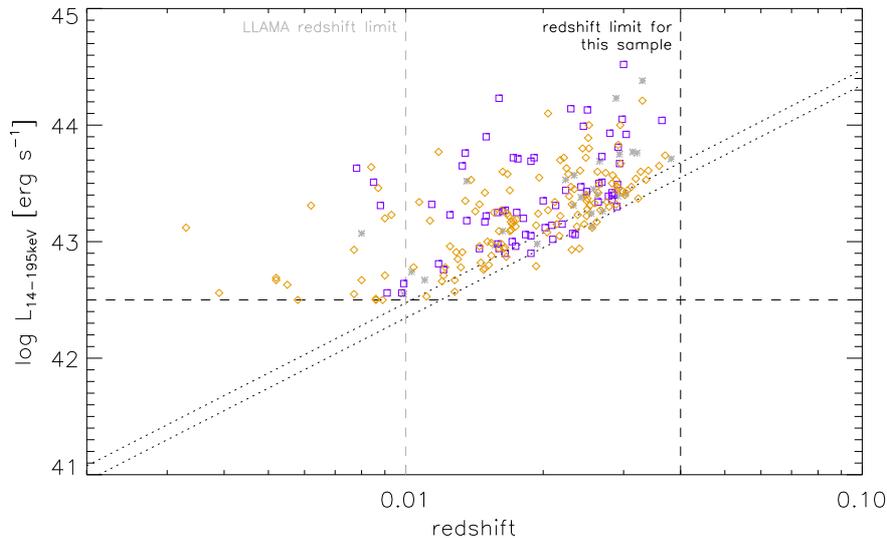}
\caption{AGN in the 70-month {\it Swift BAT} survey as for Fig.~\ref{fig:batplot}, but restricted to the 229 cross-matched with the Tully (2015a) group catalogue as described in Sec.~\ref{subsec:groupcats}.
The 67 with S0 host classifications are denoted by a purple square, the 135 with spiral hosts by an orange diamond, and the AGN with other host classifications (17 ellipticals and 10 irregulars) are marked with a grey asterisk. The distributions of the host types with redshift are shown in Fig.~\ref{fig:redshift}.}
\label{fig:batplothost}
\end{figure*}

\begin{figure}
\includegraphics[width=8cm]{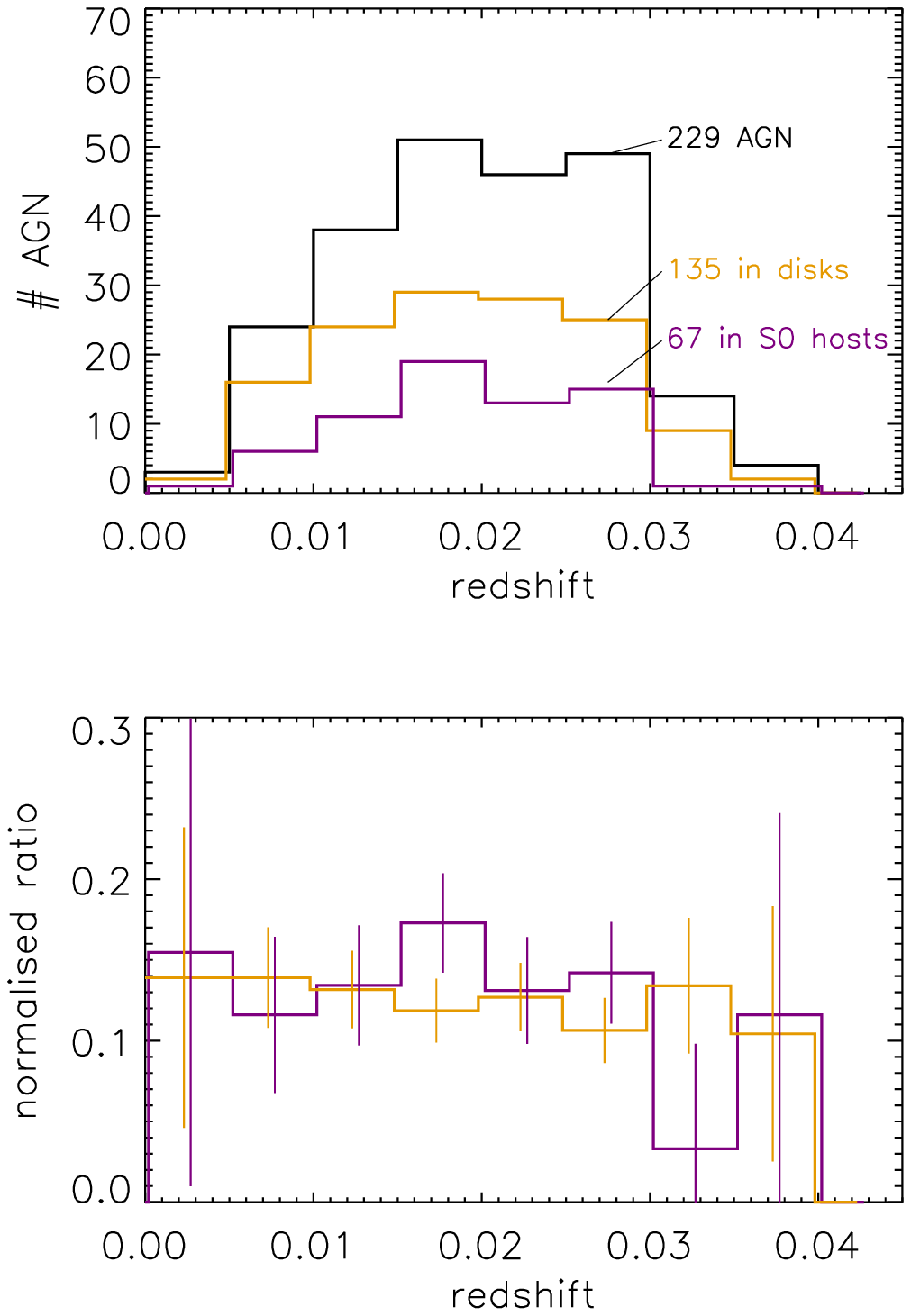}
\caption{Top panel: distribution over redshift of the AGN cross-matched with the catalogue of Tully (2015a). Additional distributions for just the AGN in spiral galaxies (orange histogram) and S0 hosts (purple histogram) are also shown.
Bottom panel: ratio of the AGN in spirals (orange) and S0 hosts (purple) to the total number of AGN at each redshift, normalised to the same mean value. Uncertainties are drawn as vertical bars on each bin. Both distributions are essentially flat, showing that there is no bias with redshift -- which could have been a concern since the sample is flux limited.}
\label{fig:redshift}
\end{figure}

\subsection{Group Catalogues}
\label{subsec:groupcats}

To assess the environment we cross-correlate the AGN with two group/cluster catalogues, which are both based largely on the 2MASS Redshift Survey \citep{huc12} of galaxies brighter than $K_s = 11.75$ but in which the groupings are defined in completely different ways. Using two independent catalogues based on the same set of galaxies allows us to minimise the impact of any bias from the way the groups are defined, and better assess the uncertainties.

The first catalogue is that of \cite{tem16}, which is based on a friends-of-friends algorithm with additional refinement procedures, applied to a combined catalogue of galaxies from several surveys of the local universe.
Since the galaxy identifications are based on the PGC catalogue (\citealt{pat03}, constituting the framework of the HYPERLEDA database), we have adopted the host galaxy morphological classifications given in the PGC catalogue.
Of the selected AGN, we find 199 in the \cite{tem16} catalogue that also have morphological host galaxy classifications.

The second catalogue is that of \cite{tul15}, for which groups were constructed using an iterative method in which initial halos are defined according to scaling relations, and then overlapping halos are assigned to a single group halo.
Host galaxy morphological classifications from \cite{vau91} are given in the data tables of this catalogue, and so we use those.
In a comparison of their catalogue to this one, \cite{tem16} pointed out that the groups are only reliable to $z\sim0.033$.
We have therefore limited our analysis with the \cite{tul15} catalogue to that redshift (noting that this restriction applies to the groups rather than individual galaxies, a few of which have redshifts greater than 0.033). This leaves 229 AGN with both group classifications and host classifications.

As comparison samples we take all the galaxies that have morphological identifications in each group catalogue to the same redshift limit.
This yields about 23500 and 27000 objects with host classifications in the \cite{tem16} and \cite{tul15} group catalogues respectively.

In order to confirm that there is no major difference in galaxy mass between the active and comparison samples, we plot in Fig.~\ref{fig:hlum} the H-band luminosity calculated from the J-H and J-K colours and K-band luminosity given by \cite{tul15}.
The H-band luminosity can be used as a proxy for stellar mass since the uncertainty in the H-band mass-to-light ratio is about 0.2\,dex (\citealt{dav15}, using masses from \citealt{kos11}).
For the AGN hosts we find a median luminosity of $\log{L_H}[L_\odot] = 10.7$ with a distribution of $\sigma = 0.3$\,dex; while for the galaxies as a whole we find $\log{L_H}[L_\odot] = 10.5$ with $\sigma = 0.4$\,dex.
When comparing the galaxy luminosities of AGN in S0 and spiral hosts, we find in both cases $\log{L_H}[L_\odot] = 10.7$ with $\sigma = 0.3$\,dex.
Using the morphological classifications in \cite{tul15} we also show in Fig.~\ref{fig:redshift} that there is no bias in host galaxy type with redshift due to the flux limited nature of the sample.

\subsection{Halo mass and Group Size}
\label{subsec:mhalo_ngrp}

\begin{figure}
\includegraphics[width=8cm]{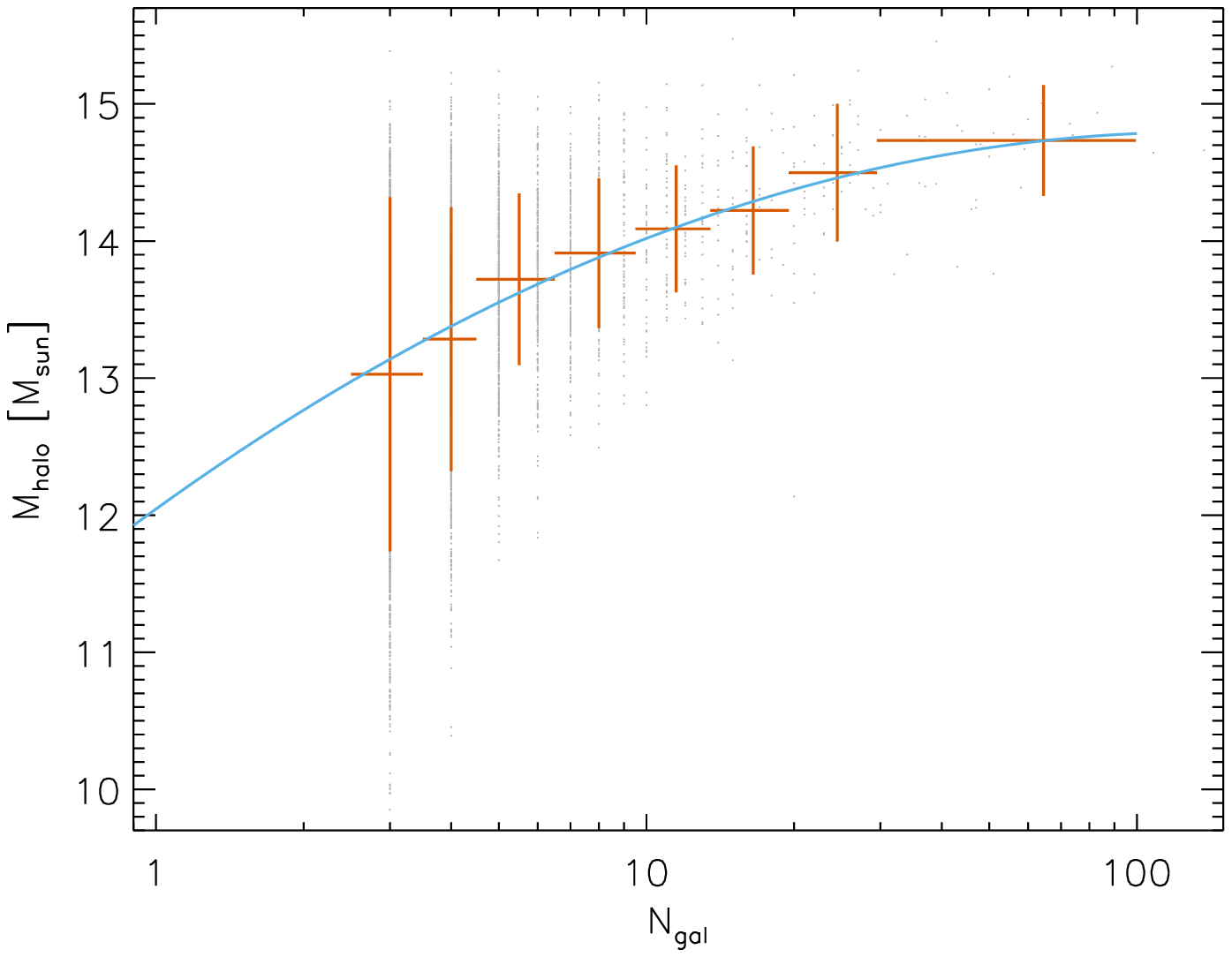}
\caption{Relation between number of galaxies in a group (halo occupation number) and halo mass, for groups in the Tempel et al. (2016) catalogue. Grey points show individual groups. These were binned together as shown by red bars, which represent the range (horizontally) for each bin and the standard deviation (vertically) about the mean of each bin.
The blue line is a quadratic fit to these binned points; and it matches the relation given in Yang et al. (2005).}
\label{fig:mhalo_ngrp}
\end{figure}

\begin{table}
\centering
\caption{Approximate relations between dark matter halo mass, the number of galaxies in the group or cluster, and the dispersion among their systemic line-of-sight velocities, from Yang et al. (2005) as well as from the data in Tempel et al. (2016), and applicable to galaxies with stellar mass of order $10^{10-11}$\,M$_\odot$.}
\label{tab:yang}
\begin{tabular}{ccc} 
\hline
log\,M$_{halo}$\,[M$_\odot$] & N$_{gal}$ & $\sigma_{los}$\,[km/s$^{-1}$] \\
\hline
12 &   1 &  90 \\
13 &   3 & 200 \\
14 &  15 & 450 \\
15 & 200 & 900 \\
\hline
\end{tabular}
\end{table}

An important consideration is whether the environment should be defined in terms of halo mass or group size (i.e. halo occupation number).
Halo mass is more closely linked to the theoretical framework of dark matter and so is easier to apply in a cosmological context, while group size is more directly linked to the observations.
Halo mass is also insensitive to the inclusion of additional low mass galaxies while the group size is dependent on the sensitivity, and hence lower mass limit, of the catalogue.
On the other hand, if using group size it is easy to distinguish between a single massive galaxy and a group of lower mass galaxies even though their respective (group) halo masses may be similar.
And, perhaps most importantly, halo mass is not easy to derive for groups with only a few galaxies.
Because most galaxies are in groups of 3 or less, we use group size as the metric for environment during our analysis.

We note, however, that there is a reasonably good relation between the group size and mean halo mass.
This is shown in Fig.~\ref{fig:mhalo_ngrp} for the groups in the \cite{tem16} catalogue, where halo mass was derived by the authors based on the kinematics and separations of the galaxies in each group.
We speculate that the large scatter in halo mass for small groups suggests that many of these groups may not be gravitationally bound (i.e. the halo mass may be over-estimated).
The figure matches a similar plot shown in Fig.~3 of \cite{yan05_358} for the groups that those authors defined using SDSS galaxies.
And similar scaling relations have been shown by \cite{wil12} and \cite{tul15iau}, the latter of which were used to build the group catalogue in \cite{tul15}.

Comparison of different results in the literature can be difficult because some authors use halo mass, while others use the dispersion of the galaxy systemic velocities or number of galaxies (halo occupation number) in the group or cluster.
An approximate conversion between these quantities is given in Table~\ref{tab:yang}.
This rough guide is valid for typical depths of relevant wide-field surveys (i.e. the depth to which all-sky surveys are largely complete within at least the local 100\,Mpc volume) over the last decade, for galaxies with stellar mass of order $10^{10}-10^{11}$\,M$_\odot$.
It is valid locally at $z=0$, but on the other hand many of the studies of halo mass of X-ray selected AGN have been performed at $z\sim1$.
\cite{mos10} looked at how the occupation number and stellar-to-halo mass ratio depend on both halo mass and redshift, finding that from $z=0$ to 1 there is at most a reduction by a factor 2 in the stellar mass for halos of $10^{12}$\,M$_\odot$, with less difference at higher halo mass.
Thus, the conversions given in Table~\ref{tab:yang} can be used up to $z\sim1$ at a precision that is sufficient for the analysis presented in this paper.

\section{The Role of Host Galaxy for Environmental Dependence}
\label{sec:z0.04}

\begin{figure*}
\includegraphics[width=6cm]{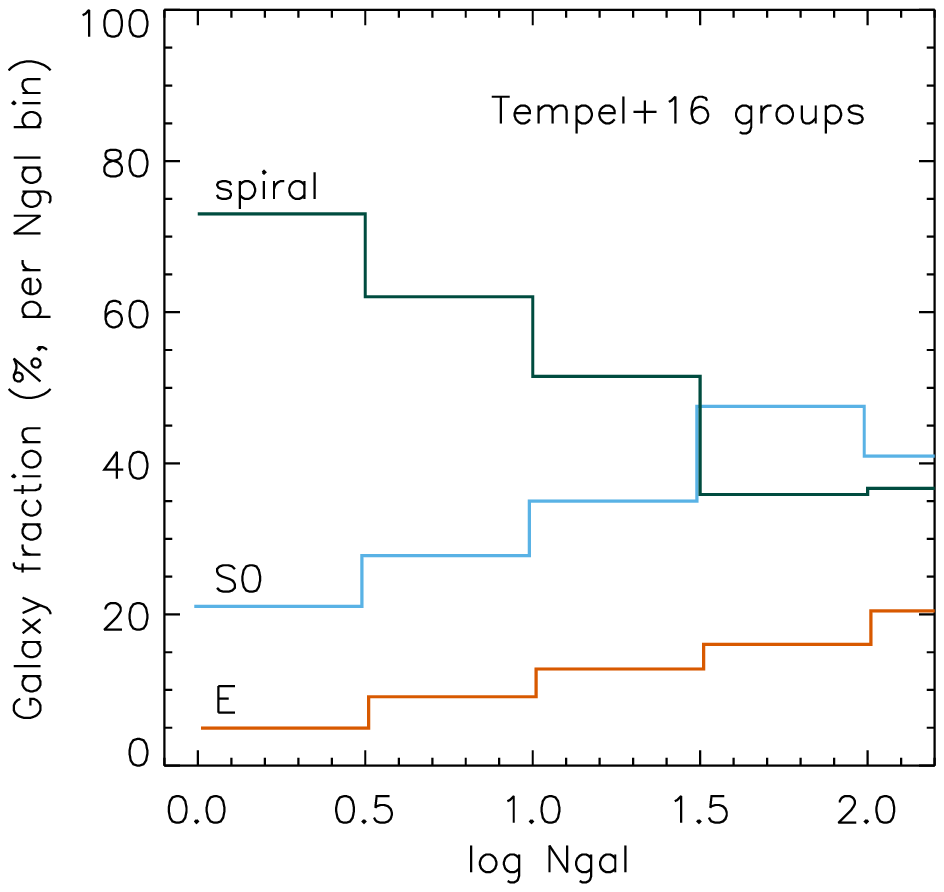}
\includegraphics[width=6cm]{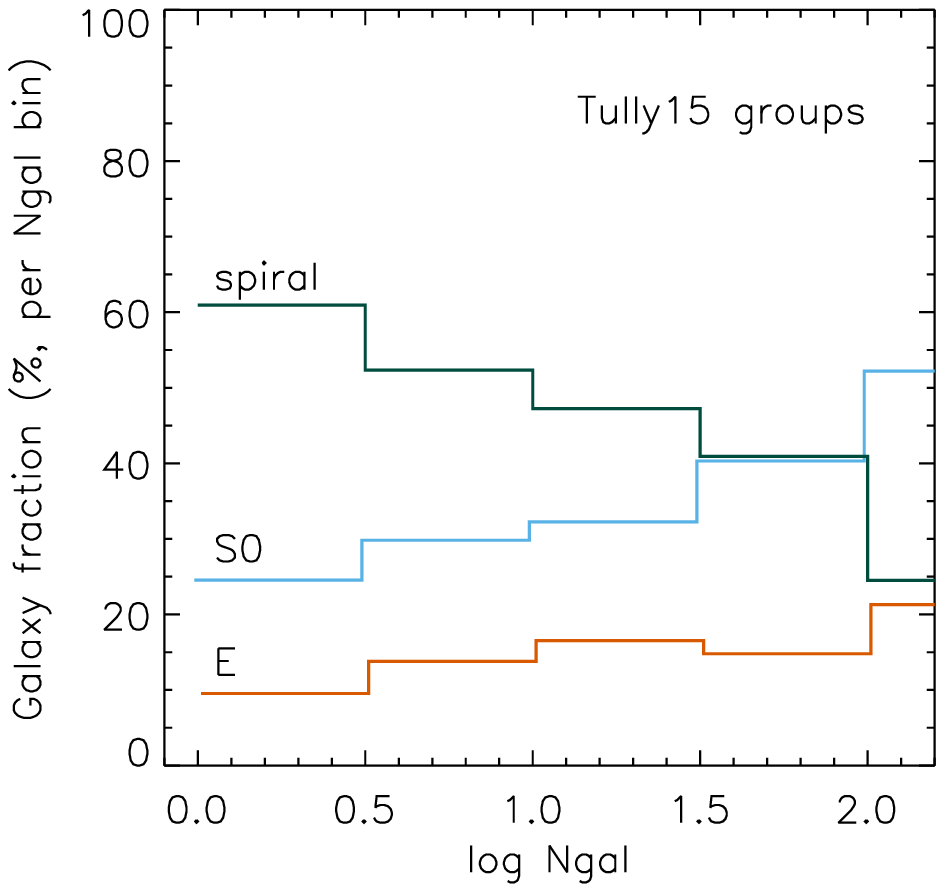}
\caption{Predominance of various host types as a function of environment; specifically, for each bin of $\log{Ngal}$, the fraction of galaxies that are spiral, S0, or elliptical.
The distributions show that the two group catalogues used in this analysis reproduce the well-established morphology density relation. Spirals (dark green) dominate in the field, but not so much in denser environments. S0 hosts (blue) become increasingly important in groups and clusters. Ellipticals are rarer, and their fraction increases with the number of galaxies in the group or cluster.}
\label{fig:distributions}
\end{figure*}

Before embarking on an analysis of the AGN environments, we first confirm that the group catalogues reproduce the well-established morphology density relation. The expected result appears in Fig~\ref{fig:distributions}, which shows that the fraction of spirals (Sa to Sd) in dense environments is reduced from about 70\% to 30\%, while the fraction of S0 (including S0/a) hosts increases from 20\% to 40\%.
And, although they are not the focus of this paper, it also shows that the fraction of ellipticals, while smaller still, also increases with environment density.
These trends have been known for decades \citep{dre80}, and have been the subject of numerous studies at redshifts from $z < 0.1$ \citep{got03,wil12} to $z \sim 1$ \citep{smi05,pos05}, and have been extended to a kinematic relation \citep{cap11}. We do not discuss this relation further, but use it only to set the context for further analysis focussing on the AGN sub-population.

\subsection{Typical Group Size and Halo Mass}
\label{subsec:typical}

\begin{figure*}
\includegraphics[width=8.5cm]{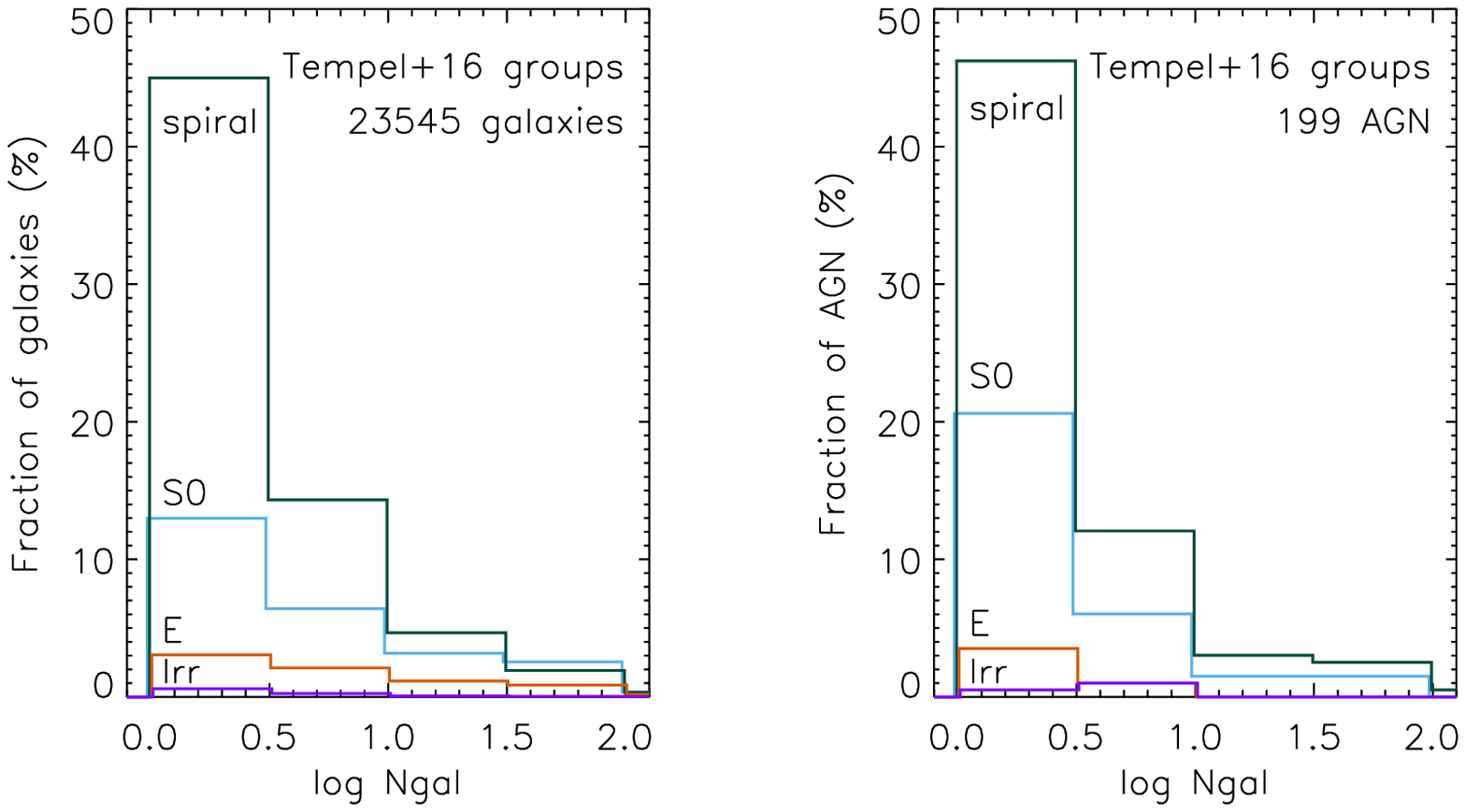}
\hspace{5mm}
\includegraphics[width=8.5cm]{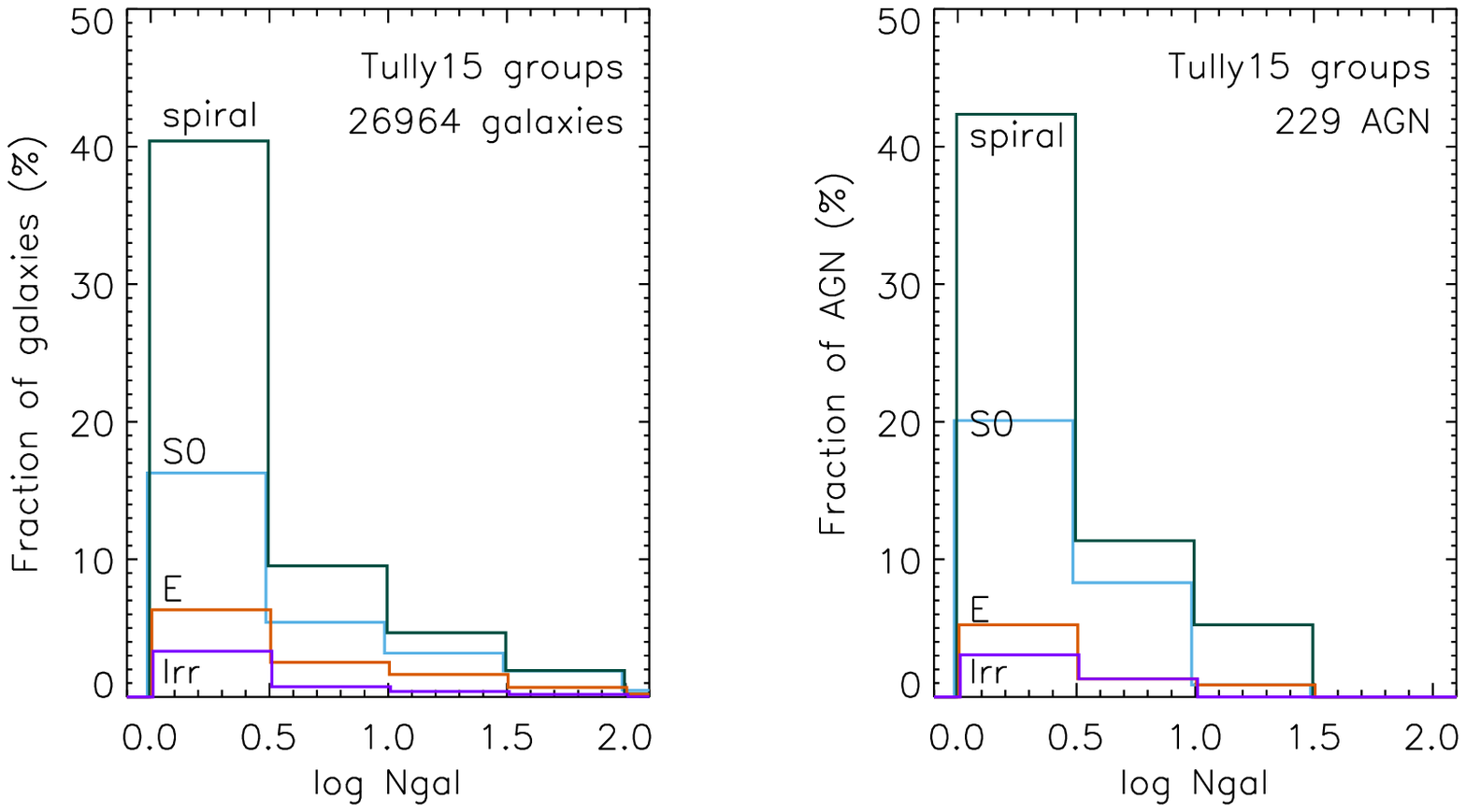}
\caption{Histogram showing how all galaxies, and the AGN host galaxies, are distributed among morphological type and group size. Results for the two group catalogues are very similar: about 2/3 of galaxies are spirals and 2/3 of hard X-ray selected AGN are in spirals; about 2/3 of galaxies and 2/3 of hard X-ray selected AGN are in groups with 3 members or less (i.e. in the field).
The predomincnace of AGN in spiral galaxies shown here, combined with the lack of an environmental dependence for those (which we show in Fig.~\ref{fig:jointplot}), dilutes any environmental dependence of the AGN population as a whole.}
\label{fig:fractions}
\end{figure*}

A number of recent studies of the environments of X-ray selected AGN have used correlation functions to assess the clustering bias on different angular scales.
They have concluded that AGN are typically found in halos with masses of $\log{M_{halo}}[M_\odot] \sim 12.5$--13 \citep{gil09,fan13,dip14,geo14}.
Based on the approximate conversion to halo occupation number given in Table~\ref{tab:yang}, this corresponds to small groups such as pairs and triplets of galaxies.
And, as can be seen in Fig.~\ref{fig:fractions}, about 2/3 of local hard X-ray selected AGN can be found in groups with 1--3 members -- a result that is consistent with the conclusions of \cite{arn09} that the fraction of X-ray selected AGN  at $0.02 < z < 0.06$ increases from clusters to groups and \cite{mar13} that at $z < 1$ the majority of AGN are found in the field.
This applies to the AGN populations as a whole, as well as to the two most common types (spiral and S0 galaxies) separately.
It also applies to the galaxy population, irrespective of whether the galaxies are active or inactive: 2/3 of the galaxies listed in the group catalogues belong to groups with only 1--3 members.
Thus we confirm the emerging consensus that the majority of AGN are found in halos containing 1--3 galaxies.
However, we caution that this is most likely due to the strongly skewed group size distribution of galaxies apparent in Fig.~\ref{fig:fractions} (i.e. completely dominated by small groups).
Unless there were a very strong trend of AGN fuelling with environment -- in the specific sense that despite the majority of galaxies being in small groups there would be almost no AGN in groups of that size -- an inevitable conclusion will be that the distribution of AGN with group size will, to zeroth order, follow that of galaxies;
and hence the typical halo/group size of AGN will roughly match that of galaxies.
Fig.~\ref{fig:fractions} shows that AGN do have a similar group size distribution as galaxies and hence appear to be distributed randomly among them.
This could at least partially be understood as a result of AGN variability \citep{hic14}.
It is only when one looks deeper, as we do in Sec.~\ref{subsec:diskbulge}, that the underlying trends and causes begin to emerge.

\subsection{AGN in disk versus bulge dominated galaxies}
\label{subsec:diskbulge}

\begin{figure*}
\includegraphics[width=12cm]{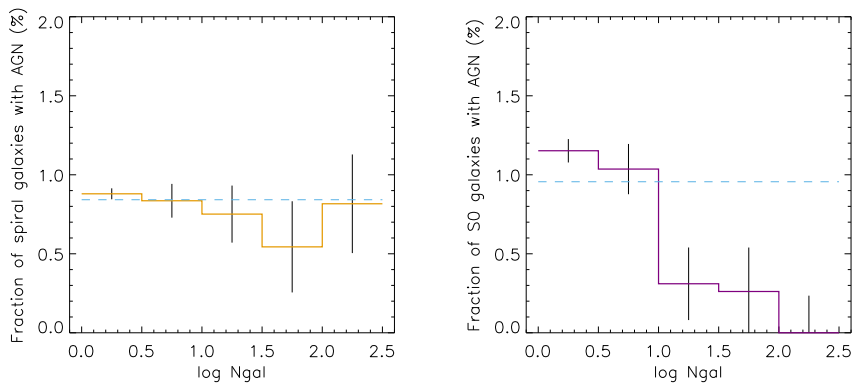}
\caption{Fraction of AGN in spiral galaxies (left) and S0 galaxies (right) as a function of the number of galaxies in the group (note that the numbers of galaxies themselves follow the distributions in Fig.~\ref{fig:fractions}). 
The histograms show the weighted average for the two group catalogues, together with the resulting 1$\sigma$ uncertainties.
The dashed blue lines show the mean for each host type. Among spiral galaxies, there is no clear evidence for an environmental dependence for AGN. In contrast, among S0 hosts, the fraction of AGN decreases strongly with environment density.}
\label{fig:jointplot}
\end{figure*}

In this section we look at the fraction of galaxies that host AGN as a function of group size (or equivalently halo mass).
We note that it is a comparative study, since the absolute AGN fractions are affected (by a factor $\sim1.5$) by the exclusion of those without morphological classifications.
Since only 5--10\% of the AGN are in ellipticals or irregulars, we focus on the two most common host types.
These are spirals (Sa to Sd), which account for almost 2/3 of the AGN; and lenticular or S0 (including S0/a) hosts, which account for almost 1/3 of the AGN.
Of the 350 hard X-ray selected AGN, we find 229 in the \cite{tul15} group catalogue that are listed with clear host morphological classifications. Of these, 67 are in S0 hosts and 135 in spiral galaxies.
Similarly, there are 199 AGN in the \cite{tem16} group catalogue with host morphological classifications, of which 59 are S0 and 128 are spiral galaxies.
Thus about 2/3 of hard X-ray selected AGN are in spirals and about 1/3 are in S0 hosts.
A similar result was reported by \cite{kos11}, who found a significant excess of spirals among {\it Swift BAT} AGN, which becomes even more pronounced at higher stellar masses.
These are both consistent with the result found by \cite{gab09} at $0.3<z<1$ in the COSMOS field, that the host galaxies of AGN span a broad range peaking between bulge-dominated and disk-dominated systems.

Using two group catalogues provides two estimates of the fraction of AGN in a given host type as a function of group size. Since the groupings are independent, we use them as if they were two separate samplings of the true distributions.
For the two host types in each catalogue, we derive the uncertainties on the fraction of AGN in each bin under the assumption that a fixed number of AGN are distributed with equal probability throughout the respective galaxy population.
We then calculate the weighted average of the two catalogues to provide the maximum likelihood estimate of the mean in each bin, together with its uncertainty.
Fig.~\ref{fig:jointplot} shows the resulting distribution, based on the two catalogues, of the AGN fraction as a function of group size.
Consistent results are also seen when looking at equivalent plots for each catalogue separately.

It is striking that the trend is different for spirals and S0 hosts.
The left panel shows no evidence for a dependence on group size of the AGN fraction in spiral galaxies (although with the data available we cannot rule out a dependence). 
Indeed, using a $\chi^2$ test, we cannot reject the null hypothesis that the data are consistent with a uniform distribution of AGN.
Given the predominance of AGN in spiral galaxies, this will tend to dilute any measurement of an environemtal impact in the population as a whole unless one takes account of the host type.
In contrast, the right panel does show a very clear trend for the AGN fraction in S0 hosts, which decreases significantly in large groups and clusters.
And a $\chi^2$ test indicates that the null hypothesis should be rejected with a 5.4$\sigma$ significance.

This result seems initially surprising because the fraction of S0 galaxies increases in denser environments.
However, it is consistent with the conclusions of \cite{dav14}.
These authors argued that if AGN in spiral galaxies are fuelled via secular processes from the gas reservoir in the host, then one can expect (i) gas to be present in both active and inactive galaxies, (ii) that the gas and stars are always co-rotating, and (iii) that there is no environmental dependence.
They tested these expectations by combining the samples of \cite{dum07} and \cite{wes12} for which there was spatially resolved kinematics for both stars and gas for active and matched inactive galaxies. 
Of the 10 AGN in disk-dominated galaxies, the gas and stars were co-rotating in all (although with misalignments up to 55$^\circ$ in some cases); and in the 7 controls, gas was detected in 5 and was also always co-rotating with the stars. Fig.~\ref{fig:jointplot} now confirms the lack of environmental dependence for AGN in spiral galaxies.

The same authors argued that if AGN in bulge-dominated galaxies were fuelled via external accretion of gas from the environment onto the galaxy, one can expect (i) a lack of gas in inactive galaxies that contrasts with the presence of gas in active galaxies, (ii) that the gas and stars should sometimes be counter-rotating, and (iii) that the environment matters.
In testing these expectations, they found that of the 11 AGN in bulge-dominated galaxies, all had gas detections; and of the 8 for which stellar kinematics could be measured, in only 5 was the gas co-rotating with the stars while 3 exhibited counter-rotation. And for the 6 controls, gas was detected in only 2. Fig~\ref{fig:jointplot} now confirms that there is an environmental dependence for AGN in S0 galaxies, and that they are preferentially found in small groups.

This conclusion is consistent with the results of surveys of early-type galaxies \citep{sar06,davt11}, which found essentially no kinematically misaligned elliptical or lenticular galaxies in clusters.
As explained also by \cite{davt16}, it implies that external accretion of cold gas is shut off in dense environments.
The reason is understood to be simply that the intra-cluster gas is ionised and so cannot easily be accreted.
That S0 galaxies cannot accrete gas in dense environments could also explain why they cannot fuel AGN in such environments.

To end this section we speculate that if, as implied above, the source of gas differs for AGN in spirals and S0 hosts, one might expect differences also in luminosity function and obscuration. These ideas are explored further in Sec.~\ref{sec:lumfunc} and~\ref{sec:obsc} respectively.
But first we attempt to verify our result in Sec.~\ref{subsec:wil12} using data from an independent analysis of AGN environment and host type.

\subsection{An Independent Comparison}
\label{subsec:wil12}

\begin{figure}
\includegraphics[width=8.5cm]{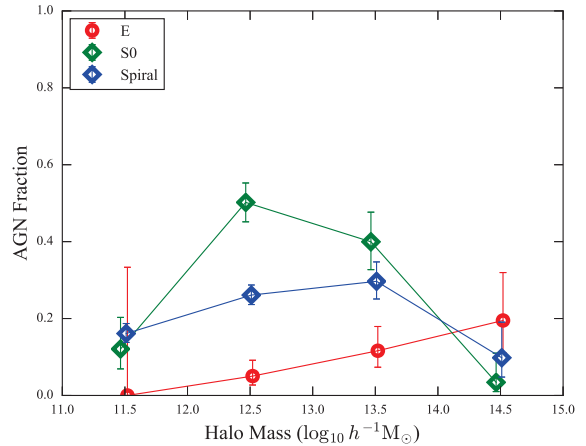}
\caption{Fraction of AGN in different types of host galaxy as a function of halo mass, from the sample of Wilman \& Erwin (2012). See Table~\ref{tab:yang} for a conversion between halo mass and group size. This shows the same result as seen in Fig.~\ref{fig:jointplot} that there is, at best, only a modest environmental dependence for AGN in spiral galaxies, but that there is a very strong decrease in the number of AGN in S0 hosts in dense environments. This shows in addition that the fraction of AGN in elliptical galaxies increases, which are mostly radio sources.}
\label{fig:wil12comp}
\end{figure}

In order to cross-check our results, we make use of an independent sample which was defined and analysed by \cite{wil12}.
It contains 911 bright ($M_B < -19$) galaxies out to $z \sim 0.04$, based on the SDSS group catalogue of \cite{yan07}.
Host types are primarily taken from \cite{vau91}, although there are some re-classifications by the authors, as well as some new classifications.
The analysis of \cite{wil12} focussed on how the morphology -- spiral, lenticular (S0), or elliptical -- relates to central versus satellite galaxies in groups and clusters; 
but it also addressed the topic of AGN, which were identified via standard optical emission line ratios.
We simply re-use their data to look at the dependence of the AGN fraction on halo mass for different host types, without distinguishing whether the galaxy is central or satellite.

\begin{figure*}
\includegraphics[width=15cm]{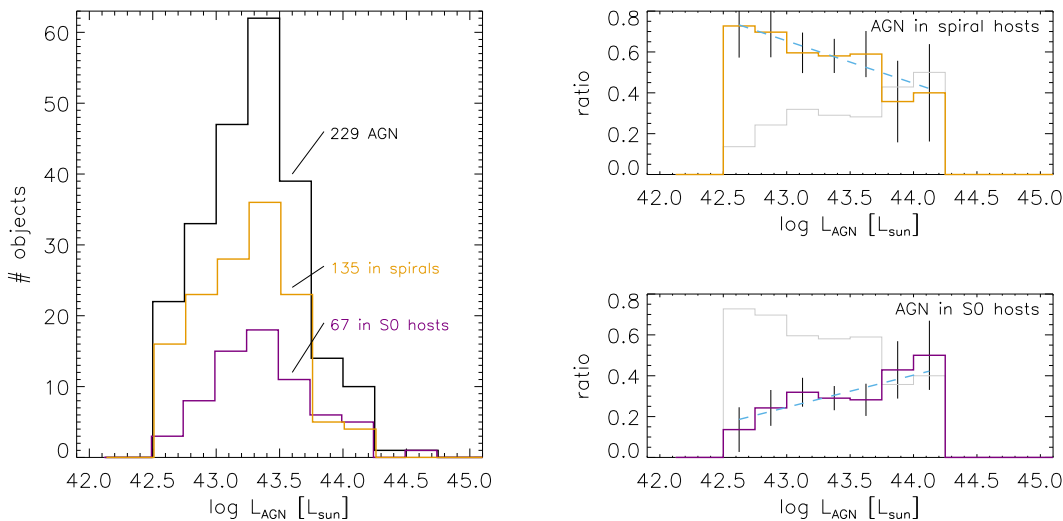}
\caption{Left: distribution of 14-195\,keV luminosity among AGN in the Tully (2015) group catalogue. Additional distributions are shown for the AGN in spiral galaxies (orange) and S0 hosts (purple). Right: ratio of the distributions for AGN in spiral (top) and S0 (bottom) galaxies to the parent distribution for all galaxies, revealing their relative luminosity functions.
The dashed blue lines show linear fit to this in each case, demonstrating their very different slopes.
In each panel, for comparison, the grey histogram traces the relative luminosity function for the other population.}
\label{fig:lumfunc}
\end{figure*}

The result is given in Fig.~\ref{fig:wil12comp}, showing the AGN fraction as a function of halo mass for elliptical, spiral, and S0 galaxies separately.
The AGN fractions, of up to a few ten percent, are very much higher than the equivalent fractions in Fig.~\ref{fig:jointplot} for our sample. The reason is that, rather than selecting only X-ray bright Seyferts, the optical line ratio selection used by \cite{wil12} includes low luminosity AGN and LINERs. 
As pointed out by \cite{ho08}, about 10\% of local galaxies have a Seyfert (and most of these are modest to low luminosity) while LINERs make up another 20\%.
As such, the difference in the numbers of AGN between our sample and that of \cite{wil12} is expected as a direct result of the AGN selection.

For AGN in S0 galaxies there is a clear peak in the AGN fraction at $\log{M_{halo}/M_\odot} = 12.5$--13.5 with a significant drop at high halo masses, while for spiral galaxies the trend is much shallower.
This is qualitatively consistent with our finding in Fig.~\ref{fig:jointplot}, that for spiral galaxies there is at most only a mild environmental dependence of AGN activity while for S0 galaxies the dependence is strong and disfavours cluster environments with large halo masses.

There is also a reduction in the AGN fractions to the smallest halo masses. We can only speculate that the reason for this may be related to the smallest halos (i.e. those below $10^{12}$\,M$_\odot$) containing only single moderate mass galaxies or small groups of low mass galaxies. One may expect these to contain AGN that, compared to those in more massive galaxies, are on average lower luminosity. And that may lead to issues about detectability, for example due to weak optical line emission.

Fig.~\ref{fig:wil12comp} also shows that the fraction of AGN in ellipticals increases with halo masses, suggesting that the large scale fuelling mechanism for these is distinct from S0 galaxies.
It is beyond the scope of this paper to discuss the cause of that trend in detail. But it is probably related to the increased incidence of elliptical galaxies in dense environments (see Fig.~\ref{fig:distributions}). Elliptical galaxies may host radio AGN, which typically accrete at lower Eddington ratios and are radiatively inefficient \citep{hec14}. They can be found in regimes where the large gas supply needed to power X-ray bright radiatively efficient AGN cannot be sustained.

\section{Hard X-ray Luminosity Function}
\label{sec:lumfunc}

The hard X-ray (14--195\,keV) luminosity distribution of the 229 AGN in our sample that are also in the \cite{tul15} group catalogue is shown as the black line in the left panel of Fig.~\ref{fig:lumfunc}.
The distributions for AGN in S0 hosts and spiral hosts are overplotted.
A simple statistical comparison using the two-sample Kolmogorov-Smirnov test shows that there is a 12\% probability that the difference between the AGN luminosities in S0 hosts and spiral hosts could arise by chance.
While this is not low enough to make the difference significant, it is important to realise that in this case the power of the test is limited because the strong central peak in the distributions is artificial.
At higher luminosities the luminosity function drops steeply, while at lower luminosities we are missing a large number of AGN due to the flux limit of the {\it Swift BAT} survey.

In order to overcome this limitation, the right panels of Fig.~\ref{fig:lumfunc} show the ratio of the distributions for AGN in spiral and S0 galaxies with respect to the parent, i.e. total AGN, distribution (and we note that Fig.~\ref{fig:redshift} implies that there should be no bias resulting from missing low luminosity AGN due to the flux limited nature of the sample).
These are relative luminosity functions in the sense that they show how the luminosity distribution for AGN in a specific type of host galaxy would appear if the luminosity distribution for the whole AGN population were flat.
Uncertainties have been derived in a similar way as described in Sec.~\ref{subsec:diskbulge}.
To do so, we have generated random subsamples, of the appropriate size for the spirals and S0s respectively, whose average luminosity distribution matches that of the parent AGN sample. 
The standard deviation of the number of sources in each bin is adopted as the uncertainty on the measurement. It should be borne in mind that an implicit assumption of this method is that the spiral and S0 subsamples have the same distribution as the parent sample. 
To assess the difference in these distributions we focus on the slope (rather than the scale, which depends only on the relative number of AGN in S0 vs spiral hosts).
The slopes for the relative luminosity distributions, shown as dashed lines in the right panels of Fig.~\ref{fig:lumfunc}, are $0.16\pm0.09$ for the AGN in S0 hosts and $-0.21\pm0.13$ for those in spirals.
A $\chi^2$ test indicates that the slopes are different with a 2.4$\sigma$ significance, confirming that the luminosities of AGN in spiral and S0 hosts are weighted towards lower and higher luminosities respectively.

We can understand this difference by considering the luminosity function as being dependent on both the black hole mass distribution and the Eddington ratio distribution.
In this perspective, it is a secondary relation (note that since it reflects the absolute accretion rate, one might be tempted to consider it a primary relation; however, there would be a second constraint of black hole mass since a black hole cannot normally accrete above its Eddington limit).
And we can show that the black hole mass distribution should differ.
\cite{lau10} (see also \citealt{wei09}) find that the bulge fraction for S0 galaxies is 0.3--0.35, decreasing to 0.25 for Sa, 0.1--0.15 for Sb to Sc, and $<$0.1 for later types.
One might therefore expect the bulge fraction to differ by a factor 2--3 between the average of the spiral galaxies and that of the S0 galaxies.
Since the galaxies in our sample have similar total masses, as apparent from Fig.~\ref{fig:hlum}, their bulge masses should differ by about the factor of 2--3 above.
Then, if these galaxies lie close to the $M_{BH}-M_{bulge}$ relation which has a slope close to 1 \citep{mar03,hae04}, their black holes may also differ by a factor of 2--4 in mass.
Hence, for similar Eddington ratios, the average luminosity of the disk-dominated galaxies is expected to be about 0.3--0.6\,dex lower than that for the bulge-dominated galaxies in our sample.
This is consistent with the 0.5\,dex difference in the median (or 0.3\,dex for the mean) luminosities of the normalised distributions.

\section{Obscuration}
\label{sec:obsc}

We can look at the fraction of obscured AGN using the Seyfert classifications given in the {\it Swift BAT} catalogue.
Before doing so, it is important to decide what counts as an obscured AGN, whether this is only type Seyfert~2, or if Seyfert~1.8--1.9 should be included, or even Seyfert~1.5?
\cite{bur15} argue that Seyfert~2 types are obscured by at least $A_V \sim 15$\,mag, and Seyfert~1i \footnote{Seyfert~1i have broad lines detected at near-infrared wavelengths but not in optical spectra. Without the near-infrared spectra they would be classified as Sy~2. But the presence of a broad line in the near-infrared indicates the obscuration is not extreme.} have $5 \la A_V \la 15$\,mag.
By deriving extinction directly from the broad lines themselves, \cite{sch16} find that Seyfert~1.8--1.9 types are obscured by $A_V \sim 5$-8\,mag, and that even Seyfert 1.5 may have $A_V \sim 3$\,mag.
These results apply to the luminosity range of the AGN analysed by \cite{sch16}, which is rather narrow with a standard deviation of 0.4\,dex around the median of $\log{L_{14-195kev}} [$erg\,s$^{-1}] = 42.7$.
In a more general context across the full range of AGN luminosities \cite{ste12} argue that much of the variation in Seyfert sub-type classification must be due to differences in the covering factor of the narrow line region, and \cite{eli14} suggest that the classification follows an evolutionary sequence.
However, the AGN in our sample also lie in a rather restricted range of moderate luminosities, and so we take the Seyfert sub-type classification as indicative of obscuration.

Several studies have looked at the question of whether dusty filaments and dust lanes along the line of sight to an AGN -- i.e. non-nuclear obscuration -- can play a role in the classification. 
In particular, \cite{pri14} showed that such phenomena can cause $A_V = 3$-6\,mag of extinction, potentially changing the AGN optical classification.
By invoking the presence of gas in the galaxy's local environment, this concept goes in the same direction as \cite{kou06}, and also earlier work by \cite{dul99}, who argue for an evolutionary scenario between Seyfert and starburst, based on the higher fraction of Seyfert~2 than Seyfert~1 galaxies with a close neighbour (they note also that there is little difference in large scale environment).
There have been a number of other studies suggesting that the local environment of Seyfert~2s is overdense compared to that of Seyfert~1s \citep{str08,jia16,gor16}.
However, these studies reach different conclusions about the scale on which the overdensity of Seyfert~2s is seen, ranging from Mpc to tens of kpc.
A more dense environment is expected to lead to more interactions and hence impact also the host galaxies, and there is some evidence for this too \citep{hun04,vil14}.
And from detailed integral field spectroscopy of molecular gas combined with dust structure maps, \cite{dav14} argue that gas and dust in the central regions of some galaxies can appear chaotic as a result of the galaxy's local environment.
Below we argue that the excess of close neighbours among Seyfert~2s may play a more important role for AGN in S0 hosts than in spiral hosts.

\begin{table}
\caption{Fractions of obscured AGN in host galaxies of different morphological types.}
\label{tab:obsc}
\begin{tabular}{lccc} 
\hline
host   & unobscured & obscured$^1$  & other$^2$ \\
galaxy & Sy~1--1.5  & Sy~1.8--2 &  \\
\hline
S0   & $33\pm6$\% ($38\pm7$\%) & $52\pm6$\% ($62\pm8$\%) & $15\pm5$\% \\
spiral & $40\pm4$\% ($48\pm5$\%) & $43\pm4$\% ($52\pm5$\%) & $18\pm3$\% \\
\hline
\end{tabular}
$^1$ We use Sy~1.8 as the threshold to define obscured AGN.\\
$^2$ `Other' refers to all classifications that are not a single Seyfert type (e.g. multiple types, ULIRG, etc). \\
Numbers in brackets refer to obscured and unobscured fractions of AGN when `other' classifications are excluded.
\end{table}

\begin{table}
\caption{Fractions of X-ray absorbed AGN in host galaxies of different morphological types.}
\label{tab:xabs}
\begin{tabular}{lcc} 
\hline
host   & unabsorbed & absorbed \\
galaxy & 
N$_H < 10^{22.3}$($10^{21.5}$)\,cm$^{-2}$ & 
N$_H \ge 10^{22.3}$($10^{21.5}$)\,cm$^{-2}$ \\
\hline
S0     & $38\pm6$\% ($33\pm6$\%) & $62\pm6$\% ($67\pm6$\%) \\
spiral & $47\pm4$\% ($39\pm4$\%) & $53\pm4$\% ($61\pm4$\%) \\
\hline
\end{tabular}
We have used threshold column of N$_H=10^{22.3}$ and N$_H=10^{21.5}$ to indicate an absorbed AGN, since these correspond to optical classifications of Sy\,2and Sy\,1.8 respectively for defining optically obscured AGN (see \citealt{bur16} and \citealt{sch16}).
\end{table}

Bearing in mind that moderate obscuration corresponding to only $A_V = 3$-6\,mag is sufficient to change the optical classification, for the purposes of this study, we consider Seyfert~1--1.5 as unobscured and Seyfert~1.8--2 to be obscured.
Table~\ref{tab:obsc} reports the fractions of obscured and unobscured Seyferts in spiral and S0 hosts.
Since some fraction of the AGN have unclear or multiple classifications (here referred to as `other'), we also show in parentheses the fraction of obscured and unobscured Seyferts excluding those.

An independent assessment of the same physical phenomenon can be found by looking at the fractions of X-ray absorbed and unabsorbed AGN in the different hosts. We have used absorbing columns derived through a consistent analysis of the 0.3--150\,keV band by Ricci et al. (in prep.) and \cite{ric15}, who provide the details for the modelling of the X-ray spectra. In our analysis we have excluded Blazars since the N$_H$ measured in these objects may be affected by the extended X-ray emission from the jet.
We have used two threshold column densities of 10$^{22.3}$\,cm$^{-2}$ and 10$^{21.5}$\,cm$^{-2}$ to define absorbed AGN, since these correspond approximately to thresholds of Sy\,2 and Sy\,1.8 in definitions of optical obscuration \citep{bur16,sch16}.
The absorbed and unabsorbed fractions are shown in Table~\ref{tab:xabs}.

Both optical obscuration and X-ray absorption show the same general trend.
For the luminosity range considered here, AGN in disk galaxies are consistent with an equal split between unobscured and obscured or equivalently unabsorbed and absorbed. 
However the fraction of obscured or absorbed AGN in S0 galaxies appears to be higher at about a 2$\sigma$ level of significance.

If AGN in S0 hosts are fuelled by externally accreted gas, then that gas could provide a source of obscuration towards the nucleus, that is additional to the gas internal to the galaxy on small scales.
Since much of the nuclear obscuration is expected to occur on small scales, on might expect that the additional obscuration on large scales should make only a minor difference to the obscured fraction.
However, this is a marginal result and so we consider it as indicative rather than robust.

\section{Conclusions}
\label{sec:conc}

We have presented an analysis of the environment, host type, and luminosity distribution for about 200 AGN selected from the {\em Swift BAT} hard X-ray survey according to $\log{L_{14-195keV}}[$erg\,s$^{-1}] > 42.5$ and $z < 0.04$.
To do so, we have used two independent group catalogues which are based on similar galaxy surveys but with the groups defined in different ways.
The main conclusions are:
\begin{itemize}
\item
Our data support the emerging consensus that the typical halo mass of local X-ray selected AGN is of order $10^{13}$\,M$_\odot$, corresponding to a typical group size of not more than a few galaxies. However, we also caution that this is most likely a consequence of the fact that most galaxies (at the sensitivity of current all-sky galaxy catalogues) are in such haloes.

\item
Most hard X-ray selected AGN are in spiral galaxies, and these show no evidence for an environmental dependence (although our data do not rule out a dependence). We argue that this is because the galaxies have their own internal gas supply which is sufficient to fuel an AGN.

\item
The fraction of S0 hosts with an AGN decreases in large groups and clusters (in contrast to the fraction of S0 galaxies itself, which, as expected for the well established morphology density relation, increases in denser environments). We argue that this is because AGN in S0 hosts are fuelled by gas from the environment that falls into the galaxy. While this is possible in small groups, gas in an intra-cluster medium is ionised and so cannot be accreted onto the galaxy efficiently.

\item
There is a difference in the luminosity functions of AGN in bulge-dominated and disk-dominated galaxies, with the latter having significantly more lower luminosity AGN. This can be understood in the context of their relative bulge sizes and the $M_{BH}-M_{bulge}$ relation.

\item
There is some evidence that the fraction of obscured AGN is higher in S0 galaxies than spirals. If confirmed, this could be due to the importance of external accretion for fuelling AGN in S0 hosts, which implies the presence of gas and dust in the group environment around the galaxy that could lead to additional obscuration.

\end{itemize}

\section*{Acknowledgements}

RD thanks all those at the Hidden Monsters conference who provided useful feedback that has been included in this paper.
R.A.R. acknowledges support from FAPERGS (project No. 2366-2551/14-0) and CNPq (project No. 470090/2013-8 and 302683/2013).





\begin{thebibliography}{99}
\bibitem[\protect\citeauthoryear{Arnold et al.}{2009}]{arn09}
Arnold T., Martini P., Mulchaey J., Berti A., Jeltema T., 2009,
ApJ, 707, 1691

\bibitem[\protect\citeauthoryear{Baumgartner et al.}{2013}]{bau13}
Baumgartner W., Tueller J., Markwardt C., Skinner G., Barthelmy S., Mushotzky R., Evans P., Gehrels N., 2013
ApJS, 207, 19

\bibitem[\protect\citeauthoryear{Burtscher et al.}{2015}]{bur15}
Burtscher L., Orban de Xivry G., Davies R., Janssen A., Lutz D., 2015,
A\&A, 578, A47

\bibitem[\protect\citeauthoryear{Burtscher et al.}{2016}]{bur16}
Burtscher L., Davies R., Graci\'a-Carpio J., Koss M., Lin M.-Y., 2016,
A\&A, 586, A28

\bibitem[\protect\citeauthoryear{Cappellari et al.}{2011}]{cap11}
Cappellari M., Emsellem E., Krajnovi\'c D., McDermid R., Serra P., et al., 2011,
MNRAS, 416, 1680

\bibitem[\protect\citeauthoryear{Davies et al.}{2014}]{dav14}
Davies R., Maciejewski W., Hicks E., Emsellem E., Erwin P., et al., 2014
ApJ, 792, 101

\bibitem[\protect\citeauthoryear{Davies et al.}{2015}]{dav15}
Davies R., Burtscher L., Rosario D., Storchi-Bergmann T., Contursi A., 2015,
ApJ, 806, 127

\bibitem[\protect\citeauthoryear{Davis et al.}{2011}]{davt11}
Davis T., Alatalo K., Sarzi M., Bureau M., Young L., et al., 2011,
MNRAS, 417, 882

\bibitem[\protect\citeauthoryear{Davis \& Bureau}{2016}]{davt16}
Davis T., Bureau M., 2016,
MNRAS, 457, 272

\bibitem[\protect\citeauthoryear{de Souza et al.}{2016}]{sou16}
de Souza R., Dantas M., Krone-Martins A., Cameron E., Coelho P., et al., 2016,
MNRAS, 461, 2115

\bibitem[\protect\citeauthoryear{de Vaucouleurs et al.}{1991}]{vau91}
de Vaucouleurs G., de Vaucouleurs A., Corwin H., et al., 1991, 
Third Reference Catalogue of Bright Galaxies (Berlin: Springer)

\bibitem[\protect\citeauthoryear{DiPompeo et al.}{2014}]{dip14}
DiPompeo M., Myers A., Hickox R., Geach J., Hainline K., 2014,
MNRAS, 442, 3443

\bibitem[\protect\citeauthoryear{Dressler}{1980}]{dre80}
Dressler A., 1980,
ApJ, 236, 351

\bibitem[\protect\citeauthoryear{Dultzin-Hacyan et al.}{1999}]{dul99}
Dultzin-Hacyan D., Krongold Y., Fuentes-Guridi I., Marziani P., 1999,
ApJ, 513, L111

\bibitem[\protect\citeauthoryear{Dumas et al.}{2007}]{dum07}
Dumas G., Mundell C., Emsellem E., Nagar N., 2007,
MNRAS, 379, 1249

\bibitem[\protect\citeauthoryear{Elitzur et al.}{2014}]{eli14}
Elitzur M., Ho L., Trump J., 2014,
MNRAS, 438, 3340

\bibitem[\protect\citeauthoryear{Fanidakis et al.}{2013}]{fan13}
Fanidakis N., Georgakakis A., Mountrichas G., Krumpe M., Baugh C., et al., 2013,
MNRAS, 435, 679

\bibitem[\protect\citeauthoryear{Gabor et al.}{2009}]{gab09}
Gabor J., Impey C., Jahnke K., Simmons B., Trump J., et al., 2009,
ApJ, 691, 705

\bibitem[\protect\citeauthoryear{Georgakakis et al.}{2014}]{geo14}
Georgakakis A., Mountrichas G., Salvato M., Rosario D., P\'erez-Gonz\'alez P., 2014,
MNRAS, 443, 3327

\bibitem[\protect\citeauthoryear{Gilli et al.}{2009}]{gil09}
Gilli R., Zamorani G., Miyaji T., Silverman J., Brusa M., et al., 2009,
A\&A, 494, 33

\bibitem[\protect\citeauthoryear{Gordon et al.}{2016}]{gor16}
Gordon Y., Owers M., Pimbblet K., Croom S., Alpaslan M., et al., 2016,
MNRAS, 2016, in press

\bibitem[\protect\citeauthoryear{Goto et al.}{2003}]{got03}
Goto T., Tamauchi C., Fujita Y., Okamura S., Sekiguchi M., Smail I., Bernardi M., Gomez P., 2003,
MNRAS, 346, 601

\bibitem[\protect\citeauthoryear{Hicks et al.}{2013}]{hic13}
Hicks E., Davies R., Maciejewski W., Emsellem E., Malkan M., Dumas G., M\"uller-S\'anchez F., Rivers A., 2013,
ApJ, 768, 107

\bibitem[\protect\citeauthoryear{H\"aring \& Rix}{2004}]{hae04}
H\"aring N., Rix H.-W., 2004,
ApJ, 604, L89


\bibitem[\protect\citeauthoryear{Huchra et al.}{2012}]{huc12}
Huchra J., Macri L., Masters K., Jarrett T., Berlind P., et al., 2012,
ApJS, 199, 26

\bibitem[\protect\citeauthoryear{Jiang et al.}{2016}]{jia16}
Jiang N., Wang H., Mo H., Dong X.-B., Wang T., Zhou H., 2016,
ApJ, 832, 111



\bibitem[\protect\citeauthoryear{Koulouridis et al.}{2006}]{kou06}
Koulouridis E., Plionis M., Chavushyan V., Dultzin-Hacyan D., Krongold Y., Goudis C., 2006,
ApJ, 639, 37

\bibitem[\protect\citeauthoryear{Heckman \& Best}{2014}]{hec14}
Heckman T., Best P., 2014,
ARA\&A, 52, 589

\bibitem[\protect\citeauthoryear{Hickox et al.}{2014}]{hic14}
Hickox R., Mullaney J., Alexander D., Chen C.-T., Civano F., et al., 2014
ApJ, 782, 9

\bibitem[\protect\citeauthoryear{Ho}{2008}]{ho08}
Ho L., 2008,
ARA\&A, 46, 475

\bibitem[\protect\citeauthoryear{Hunt \& Malkan}{2004}]{hun04}
Hunt L., Malkan M., 2004,
ApJ, 616, 707

\bibitem[\protect\citeauthoryear{Koss et al.}{2010}]{kos10}
Koss M., Mushotzky R., Veilleux S., Winter L., 2010
ApJL, 716, L125

\bibitem[\protect\citeauthoryear{Koss et al.}{2011}]{kos11}
Koss M., Mushotzky R., Veilleux S., et al., 2011
ApJ, 739, 57

\bibitem[\protect\citeauthoryear{Laurikainen et al.}{2010}]{lau10}
Laurikainen E., Salo H., Buta R., Knapen J., Comer\'on S., 2010,
MNRAS, 405, 1089

\bibitem[\protect\citeauthoryear{Li et al.}{2006}]{li06}
Li C., Kauffmann G., Wang L., White S., Heckman T., Jing Y., 2006,
MNRAS, 373, 457

\bibitem[\protect\citeauthoryear{Marconi \& Hunt}{2003}]{mar03}
Marconi A., Hunt L., 2003,
ApJ, 589, L21

\bibitem[\protect\citeauthoryear{Martini et al.}{2013}]{mar13}
Martini P., Miller E., Brodwin M., Stanford S., Gonzalez A., 2013,
ApJ, 768, 1

\bibitem[\protect\citeauthoryear{Moster et al.}{2010}]{mos10}
Moster B., Somerville R., Maulbetsch F., van den Bosch F., Macci\'o A., Naab T., Oser L., 2010
ApJ, 710, 903

\bibitem[\protect\citeauthoryear{M\"uller-S\'anchez et al.}{2013}]{mue13}
M\"uller-S\'anchez F., Prieto M.A., Mezcua M., Davies R., Malkan M., Elitzur M., 2013,
ApJ, 763, L1

\bibitem[\protect\citeauthoryear{Paturel et al.}{2003}]{pat03}
Paturel G., Petit C., Prugniel P., Theureau G., Rousseau J., Brouty M., Dubois P., Cambresy L., 2003,
A\&A, 412, 45

\bibitem[\protect\citeauthoryear{Postman et al.}{2005}]{pos05}
Postman M., Franx M., Cross N., Holden B., Ford H., et al., 2005,
ApJ, 623, 721

\bibitem[\protect\citeauthoryear{Prieto et al.}{2014}]{pri14}
Prieto M.A., Mezcua M., Fern\'andez-Ontiveros J., Schartmann M., 2014,
MNRAS, 442, 2145

\bibitem[\protect\citeauthoryear{Ricci et al.}{2015}]{ric15}
Ricci C., Ueda Y., Koss M., Trakhtenbrot B., Bauer F., Gandhi P., 2015
ApJ, 815, L13

\bibitem[\protect\citeauthoryear{Sabater et al.}{2013}]{sab13}
Sabater J., Best P., Argudo-Fern\'andez M., 2013,
MNRAS, 430, 638

\bibitem[\protect\citeauthoryear{Sabater et al.}{2015}]{sab15}
Sabater J., Best P., Heckman T., 2015,
MNRAS, 447, 110

\bibitem[\protect\citeauthoryear{Sarzi et al.}{2006}]{sar06}
Sarzi M., Falc\'on-Barroso J., Davies R., Bacon R., Bureau M., et al., 2006,
MNRAS, 366, 1151

\bibitem[\protect\citeauthoryear{Schade, Boyle, \& Letwasky}{2000}]{sch00}
Schade D., Boyle B., Letawsky M., 2000
MNRAS, 315, 498

\bibitem[\protect\citeauthoryear{Schnorr-M\"uller et al.}{2016}]{sch16}
Schnorr-M\"uller A., Davies R., Korista K., Burtscher L., Rosario D., et al., 2016,
MNRAS, 462, 3570

\bibitem[\protect\citeauthoryear{Serber et al.}{2006}]{ser06}
Serber W., Bahcall N., M\'enard B., Richards G., 2006,
ApJ, 643, 68

\bibitem[\protect\citeauthoryear{Smith et al.}{2005}]{smi05}
Smith G., Treu T., Ellis R., Moran S., Dressler A., 2005,
ApJ, 620, 78

\bibitem[\protect\citeauthoryear{Stern \& Laor}{2012}]{ste12}
Stern J., Laor A., 2012,
MNRAS, 426, 2703

\bibitem[\protect\citeauthoryear{Strand, Brunner, \& Myers}{2008}]{str08}
Strand N., Brunner R., Myers A., 2008,
ApJ, 688, 180

\bibitem[\protect\citeauthoryear{Tempel et al.}{2016}]{tem16}
Tempel E., Kipper R., Tamm A., Gramann M., Einsato M., Sepp T., Tuvikene T., 2016, 
A\&A, 588, 14

\bibitem[\protect\citeauthoryear{Tully}{2015a}]{tul15}
Tully R.B., 2015,
AJ, 149, 171

\bibitem[\protect\citeauthoryear{Tully}{2015b}]{tul15iau}
Tully R.B., 2015,
In {\it Scale-Free Processes in the Universe}, invited talk given at IAU General Assembly XXIX, Focus Meeting 18; arXiv 1508.05934

\bibitem[\protect\citeauthoryear{Villaroel \& Korn}{2014}]{vil14}
Villarroel B., Korn A., 2014, 
Nature Physics, 10, 417

\bibitem[\protect\citeauthoryear{Weinzirl et al.}{2009}]{wei09}
Weinzirl T., Jogee S., Khochfar S., Burkert A., Kormendy J.,
ApJ, 696, 411

\bibitem[\protect\citeauthoryear{Westoby et al.}{2012}]{wes12}
Westoby P., Mundell C., Nagar N., Maciejewski W., Emsellem E., Roth M., Gerssen J., Baldry I., 2012,
ApJS, 199, 1

\bibitem[\protect\citeauthoryear{Wilman \& Erwin}{2012}]{wil12}
Wilman D., Erwin P., 2012,
ApJ, 746, 160

\bibitem[\protect\citeauthoryear{Winter et al.}{2012}]{win12}
Winter L., Veilleux S., McKernan B., Kallman T., 2012,
ApJ, 745, 107

\bibitem[\protect\citeauthoryear{Yang et al.}{2005}]{yan05_358}
Yang X., Mo H., Jing Y., van den Bosch F., et al., 2005,
MNRAS, 358, 217

\bibitem[\protect\citeauthoryear{Yang et al.}{2007}]{yan07}
Yang X., Mo H., van den Bosch F., et al., 2007,
ApJ, 671, 153

\end{thebibliography}








\bsp	
\label{lastpage}
\end{document}